%% file: DIS01_LANL.tex
\documentclass{ws-p8-50x6-00}
\usepackage[english]{babel}
\begin{document}
\include{DIS01_LANL-txt}

%
%
\end{document}

%% file: DIS01_LANL-txt.tex
\title{
ZEUS at HERA II
}                                                       
                    
\author{B. Foster}
\address{H.H. Wills Physics Lab., Tyndall Avenue, Bristol, BS8 1TL, U.K. \\ 
E-mail: b.foster@bris.ac.uk}

\date{15.\ May 2000}

\maketitle

\abstracts{The ZEUS detector has been upgraded in a number of areas to prepare for the physics opportunities of HERA II. These upgrades, and their
physics rationale and promise, are briefly outlined. The measurement
of polarisation at HERA II, and its importance for the HERA II physics
programme, is also discussed.}

\section{Introduction}
\label{sec-int}

The HERA upgrade will produce about a factor five improvement in luminosity
delivered to the experiments and aims to accumulate 1 fb$^{-1}$ of data in the
HERA II programme. In order to take advantage of this, 
the ZEUS detector has been
upgraded in several areas: 
the silicon microvertex detector (MVD); the straw-tube tracker (STT); and the luminosity monitor. 
The addition of spin rotators will produce longitudinal polarisation in
both ZEUS and H1; both experiments have collaborated with HERMES and 
HERA to produce an accurate measurement of the
degree of polarisation. 

This talk reviews the characteristics and status of the
major upgrades, together with the physics towards which they are
aimed. It is not a comprehensive review of HERA II physics, and should
be read in conjunction with the talks at this workshop by E. Elsen,
which covers complementary areas, and by W. Buchm\"{u}ller.

\section{The vertex region}
\label{sec-vert}

\begin{figure}[t]
\epsfxsize=25pc 
\epsfbox{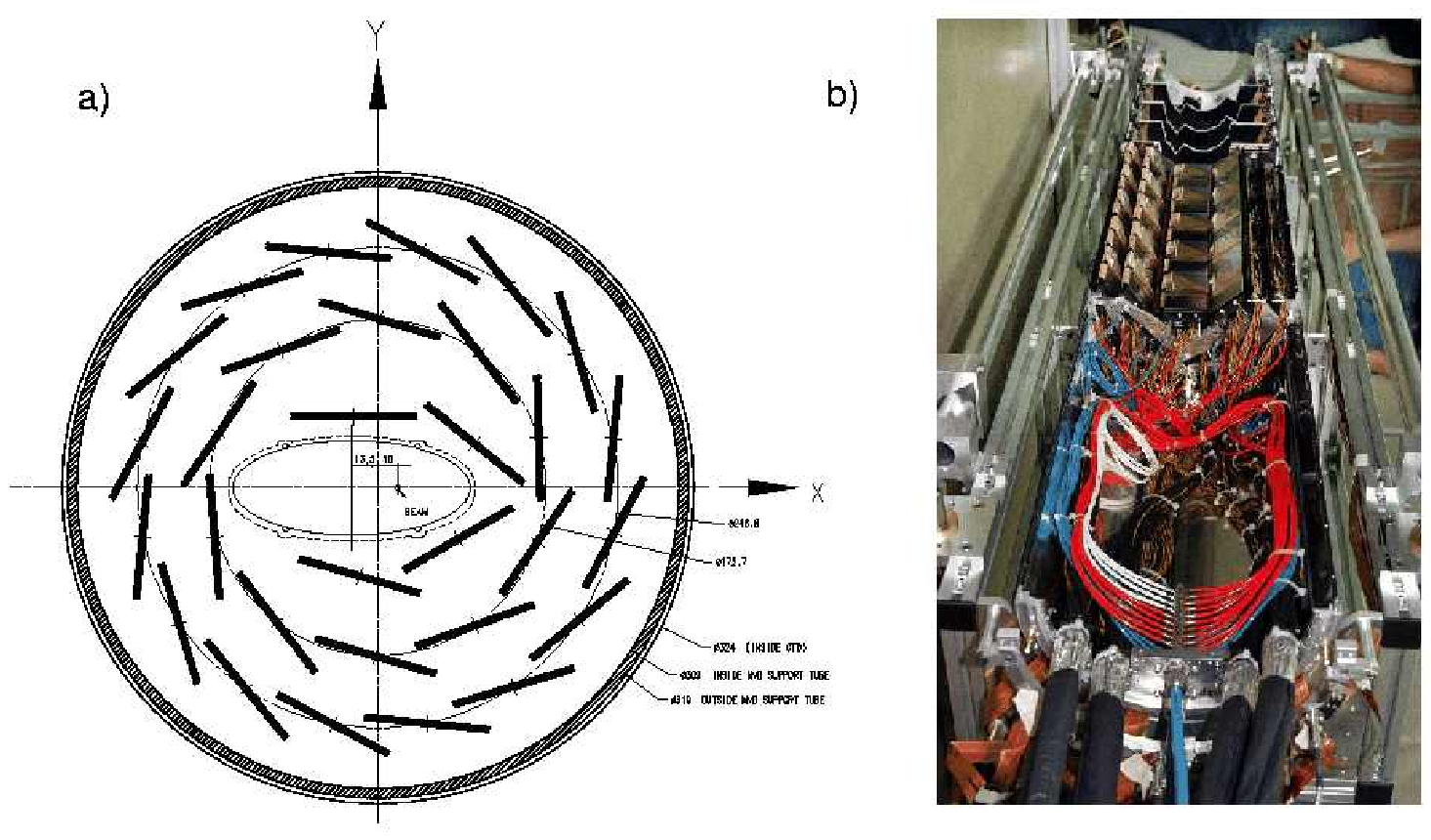} 
\caption{a) A section through the barrel MVD, showing the arrangement around the beam-pipe of each of the MVD ladders. b) A photograph of one half of the MVD,
showing the barrel ladders, one half of each of the four forward ``wheels'' and
the cables and services.
}
\label{fig:MVD-barrel-photo}
\end{figure}

The ZEUS MVD~\cite{nim:a435:34} 
consists of 20 $\mu$m pitch $n$-type silicon-strip detectors 
with $p^+$-type implants. The readout pitch is 120 $\mu$m, leading
to more than 200,000 readout channels, which are digitised by a
custom-built clock, control and ADC system. 
The detectors are organised in two main groups: a ``barrel'', which surrounds the elliptical 2 mm-thick ($\sim 1.1$\% of a radiation length) 
aluminium-beryllium beam-pipe; and four ``wheels'', consisting
of wedge-shaped detectors mounted 
perpendicular to the beam-line and displaced in the proton, or ``forward'' direction. The layout of the barrel detector is shown
in Fig.~\ref{fig:MVD-barrel-photo}a); it 
gives maximum coverage for charged particles emanating from
the interaction point while accommodating the elliptical beam-pipe. 
Most tracks pass through three separate detectors as they leave
the interaction point; the remainder pass through two. Figure~\ref{fig:MVD-barrel-photo}b) shows one half
of the MVD before installation at DESY. In the barrel region, the ladders, each of which consists of 5 silicon detectors, and halves of the
four forward ``wheels'', can be seen, as can the dense array of readout and
services cables and the cooling system. The complete MVD was installed in ZEUS
in April 2001. 

\begin{figure}[h]
\epsfxsize=25pc 
\epsfbox{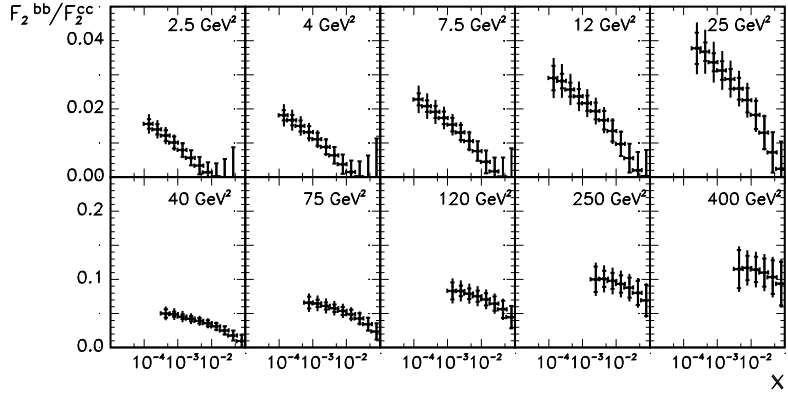} 
\caption{The ratio of the contribution of $b$-quark to $c$-quark production
in $Q^2$ bins as a function of Bjorken $x$.
}
\label{fig:F2b-c-ratio}
\end{figure}

The physics programme addressed by the MVD is that of the flavour decomposition of the proton and photon and the search for physics beyond the Standard Model.
At HERA I, precise measurements of the inclusive structure function, $F_2$,
have been made over a large kinematic range. In
addition, measurements of the semi-inclusive charm structure function, $F_2^c$, have also been made, but these are limited by
statistics. The large increase in luminosity of HERA II, 
together with the ability to
tag heavy-quark decays in the MVD, should improved 
the measurement of $F_2^{c}$. After about 500 pb$^{-1}$, 
an uncertainty of around the 2\% currently measured on $F_2$ should
be obtained. In addition, $b$-quark production can be measured precisely; a
Monte Carlo simulation~\cite{proc:hera:1995:89} of a measurement of
$F_2^{b}/F_2^c$ after 500 
pb$^{-1}$ is shown in 
Figure~\ref{fig:F2b-c-ratio}. 

It should also be possible, from a combination on neutral and charged current
measurements, to separate out the $u,d,s,c,b$ and $g$ contribution to $F_2$. The
MVD will not only be vital in the extraction of the $b$ and $c$ signal; it
will also be important to measure the strange quark contribution~\cite{proc:hera:1995:102}, by looking
for charm in charged current interactions via the transition $s \stackrel{W}{\rightarrow} c$. Since many possible exotic particles tend
to decay preferentially to heavy quarks, the MVD will also help in searches for
physics beyond the Standard Model. 

\section{Charged-particle tracking in the forward direction}
\label{sec-stt}

\begin{figure}[h]
\epsfxsize=28pc 
\epsfbox{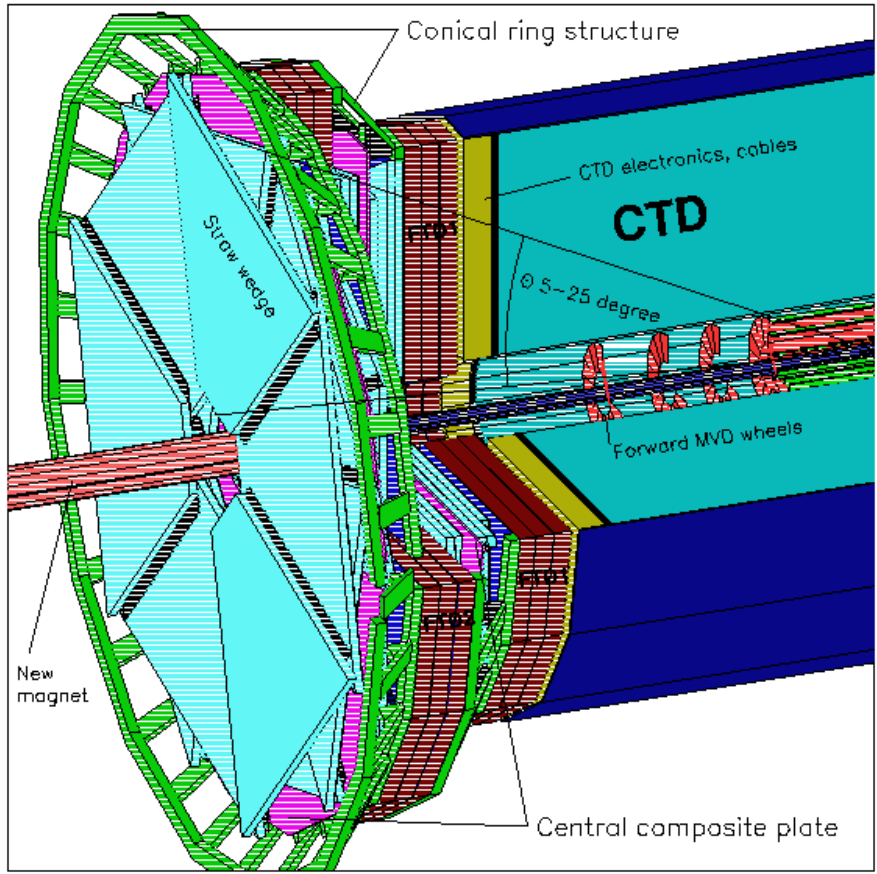} 
\caption{Schematic view of the two STT supermodules and two
of the three planar drift chamber modules. 
}
\label{fig:STT}
\end{figure}

\begin{figure}[h]
\epsfxsize=23pc 
\epsfbox{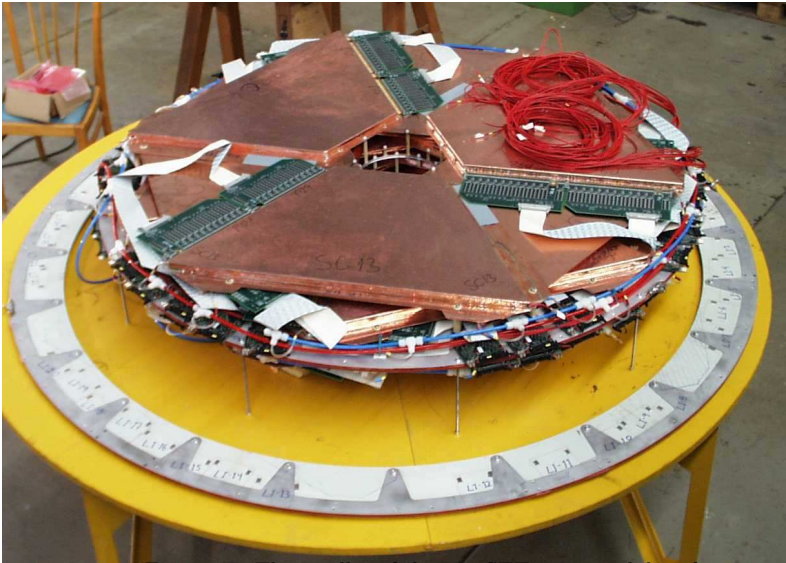} 
\caption{\vspace*{2cm} The smaller of the two STT supermodules after assembly.
}
\label{fig:STT_photo}
\end{figure}

The higher luminosity expected at HERA II will increase the number of very
high $Q^2$ events in which the electron or positron is scattered into the forward direction. It will also give access to rare processes, including
possible physics beyond the Standard Model, which tend to have 
forward jets. The pattern-recognition capabilities of the ZEUS Forward
Tracker have therefore been improved by the replacement of two layers of transition radiation detector by layers of straw tubes, as shown schematically
in Fig.~\ref{fig:STT}. The straws are approximately 7.5 mm in diameter and
range in length from around 20 cm to just over 1 m. They are constructed from
two layers of 50 $\mu$m kapton foil coated with a 0.2 $\mu$m layer of aluminium, surrounding a 50 $\mu$m wire
at the centre. The straws are arranged in
wedges consisting of three layers rotated with respect to each other to give three-dimensional reconstruction. Each of the two `supermodules' consists of four layers of such wedges. Figure~\ref{fig:STT} shows, schematically, the arrangement of the new forward detector, together with the central tracking
detector (CTD) and the MVD. The smaller of the two supermodules is
shown after construction in Figure~\ref{fig:STT_photo}. 

Groups of six straws are multiplexed to a single FADC 
channel. New front-end pre-amplification and shaping circuitry based on the ASDQ chip, as well as cable drivers, were constructed. The detector was
installed in ZEUS in May, 2001.

\begin{figure}[h]
\epsfxsize=28pc 
\epsfbox{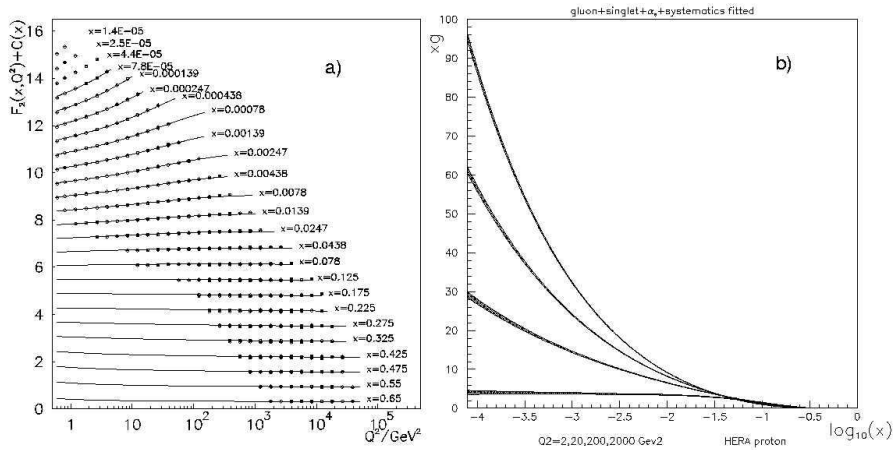} 
\caption{a) Estimate of the accuracy with which $F_2$ can be measured
in $x$ bins as a function of $Q^2$ with 1 fb$^{-1}$ at HERA II. b) Estimate
of the accuracy to which the gluon distribution function can be extracted from
a QCD fit to such a data sample.
}
\label{fig:F2}
\end{figure}

One of the major gains in the improved forward tracking will be
in the measurement of the $F_2$ structure function at high $Q^2$
and $x$. An estimate~\cite{proc:hera:1995:33} of the accuracy that can be achieved with 1 fb$^{-1}$ is shown in Fig.~\ref{fig:F2}, together
with the corresponding accuracy of the gluon distribution function
extracted from a QCD fit to this data. 

\section{Luminosity monitor}
\label{sec-lumi}

\begin{figure}[h]
\epsfxsize=28pc 
\epsfbox{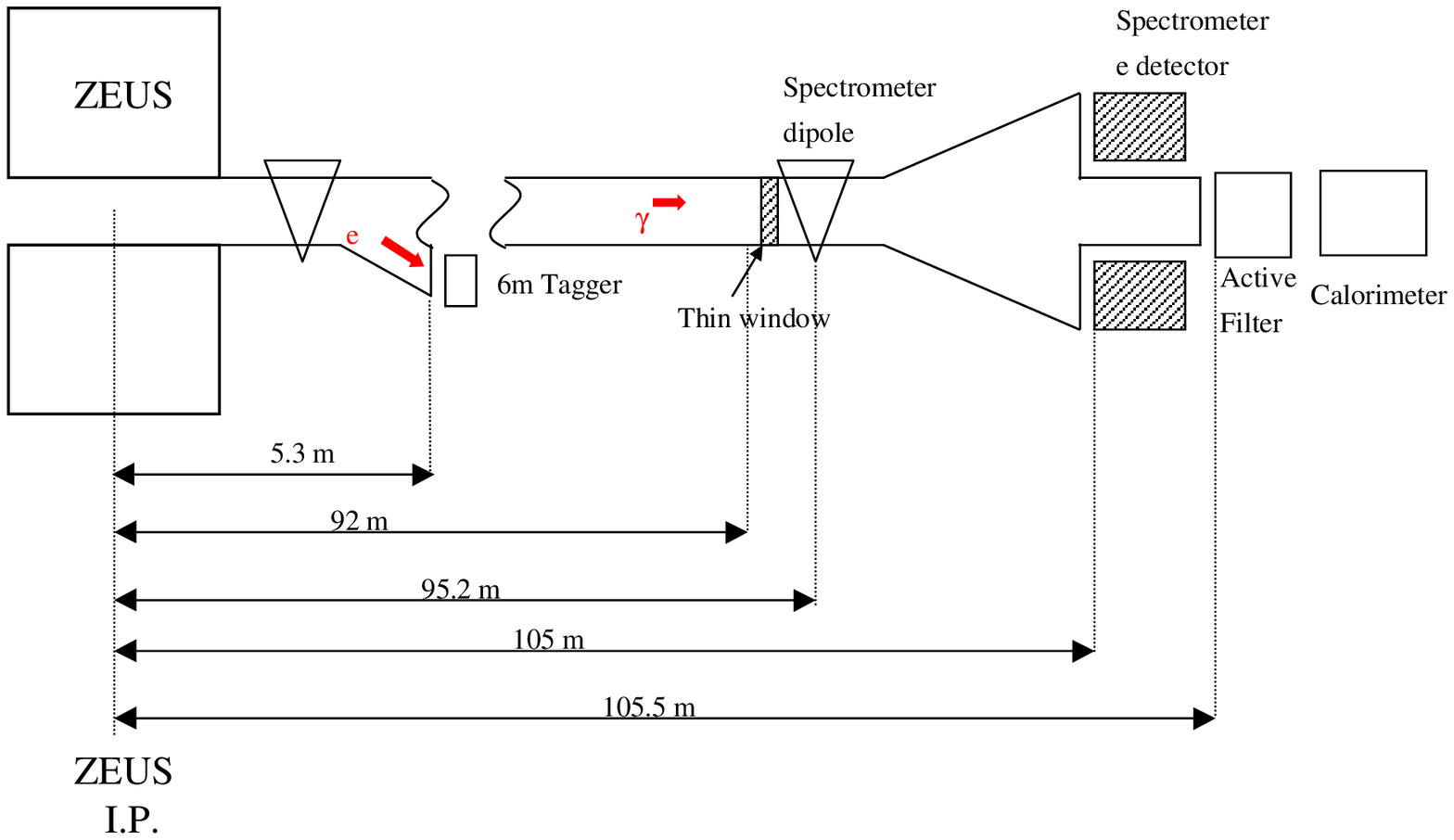} 
\caption{Schematic view of the ZEUS luminosity monitor system,
consisting of an electron tagger, a pair spectrometer, and a 
photon calorimeter with active filter.
}
\label{fig:lumischem}
\end{figure}

The high luminosity at HERA II together with the large synchrotron-radiation
background require a new approach to the measurement of luminosity at ZEUS.
The simple photon calorimeter used previously must be upgraded to cope with the background and the increased probability for multiple bremsstrahlung photons in one beam-crossing. In
order to improve the accuracy of the luminosity measurement, a second technique,
with very different systematic uncertainties, based on an electron-positron
pair spectrometer, has been constructed. Both devices use the information 
from a small calorimeter placed around 6 m from the interaction point which 
detects the radiating electron. The set-up is shown schematically in Fig.~\ref{fig:lumischem}.

The photon calorimeter is a lead-scintillator sandwich with
a position detector consisting of strips of scintillator. 
In order to cope with the synchrotron radiation background, 
an ``active filter'', consisting of
two carbon absorbers, each of two radiation lengths, alternating with Aerogel Cerenkov detectors has been constructed. The absorbers protect the calorimeter
from radiation damage, while the Cerenkov detectors detect 
high-energy photons that convert in the absorbers, allowing
the calorimeter energy to be corrected and good resolution to be
recovered. The calorimeter is currently being installed in the HERA tunnel.

The pair spectrometer is situated downstream of an exit window 
corresponding to around 12\% of a radiation length. The
electron-positron pairs that covert therein are separated by a dipole magnet
and detected in a pair of tungsten-scintillator sandwich calorimeters,
previously used to detect small-angle, very low-$Q^2$ electrons close
to the beam-line in the main ZEUS detector. These devices have been refurbished
and installed in the spectrometer in April 2001.  

The ``6 m tagger'' consists of a $10 \times 10 \times 5$ cm tungsten-scintillating fibre calorimeter next to the beam-pipe inside
one of the HERA magnets. The device has been 
tested in the DESY test beam and shown to be linear
to better than 1\%, to have a relative energy resolution of 
17\%/$\sqrt{E}$ and a uniformity better than 5\%. In order to protect
against possible radiation damage, the tagger will be
installed after HERA commissioning is complete.

Each of these devices uses a common electronic readout system. This
will be completed, installed and tested in the next few months. It is hoped
to attain a luminosity measurement precision of around 1\%. 

\section{Polarisation}
\label{sec-pol}

For several
years, HERMES have trail-blazed the use
of polarised electrons at HERA 
and, in close partnership with the
HERA accelerator physicists, polarisations of around 65\% have been
achieved. It is hoped to increase the
accuracy with which the polarisation can be measured to $\delta P/P \sim$ 2\%
per bunch per minute. 
This will be achieved by a collaboration between H1, HERMES,
ZEUS and the HERA machine in the POL2000 project. The collaboration
is constructing two instruments, both of which detect
asymmetries in back-scattered light from high-intensity polarised lasers. One,
the LPOL, measures the energy asymmetry between left- and right-handed polarised
scattered photons, which, in the photon-electron centre-of-mass frame, can
be understood as arising from 
the relative orientation of the electron and photon spin leading to orbital angular momentum of either 1/2 or 3/2. When
boosted back to the laboratory system, this results in a difference
in energy between the two situations. It is proposed to use a Fabry-P\'{e}rot cavity in order
to permit very fast bunch-by-bunch measurements. The other detector, the TPOL, measures the degree of transverse polarisation by measuring the up-down asymmetry of the detected photons. This asymmetry arises since the transverse spin direction defines a preferred spatial axis with respect to the beam direction. The TPOL uses a high-precision position detector, which is
the ZEUS contribution to the POL2000 project, plus a calorimeter to
measure this asymmetry. 

\begin{figure}[t]
\epsfxsize=28pc 
\epsfbox{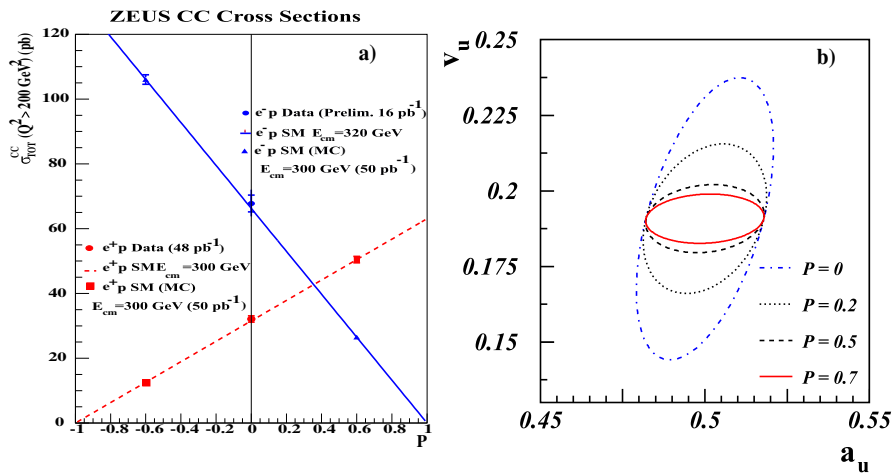} 
\caption{a) The cross section for charged current interactions. The points at
P=0 are obtained from ZEUS preliminary results at the indicated centre-of-mass
energies, while those at non-zero polarisation are Monte Carlo simulations of the expected accuracy in ZEUS assuming the Standard Model cross section for an integrated luminosity of 50 pb$^{-1}$ per point. b) One standard-deviation
contours for the $u$-quark vector and axial couplings from fits to Monte Carlo data in which the $d$-quark couplings are held constant. The various contours
correspond to 1 fb${-1}$ of data with the degrees of polarisation indicated.
}
\label{fig:e+-pol}
\end{figure}

The combination of high-precision measurements of both luminosity and
polarisation will be important in a wide range of HERA II physics.
The most obvious application is in the electroweak sector, where, 
for example, the charged current cross section should vanish for the appropriate
combinations of lepton charge and polarisation. A measurement at three
polarisations, such as shown in Fig.~\ref{fig:e+-pol}a), even
with relatively modest luminosity, will provide
an accurate test of this prediction, and provide mass limits on right-handed
$W$s higher than 400 GeV. 

Strong polarisation effects are also predicted at high $Q^2$ in
the neutral current, where, e.g. at $Q^2 = 10^4$ GeV$^2$ and 
$x = 0.2$, there is a factor two difference between the 
predicted cross sections for left- and right-handed electrons.
The luminosity available at HERA II will permit the determination
of the $u$- and $d$-quark couplings. The gain in the determination
of the $u$-quark electroweak couplings from 1 fb$^{-1}$ of simulated
data~\cite{proc:hera:1995:163} as the degree of polarisation
is increased is clearly visible in 
Fig.~\ref{fig:e+-pol}b). 

In addition to the use of precise luminosity and polarisation information
in the study of electroweak processes, polarisation also offers an invaluable
tool in the study of possible signals beyond the Standard Model. Varying the polarisation to reduce the
cross sections of Standard Model processes could well improve the
signal to background for new physics signals, such as leptoquarks or supersymmetric particles that violate $R$ parity, for which
HERA will still have the highest sensitivity for the next few years. 

\section{Summary}
\label{sec-summary}
The summer of 2001 will see HERA and the general-purpose experiments H1 and ZEUS upgraded and operational. The promise of the HERA II programme is great. HERA I produced deep insights into QCD and laid the foundations for the study
of the space-like electroweak interaction at high $Q^2$. HERA II promises
to build on those foundations to open new fields of precision
electroweak study and searches for physics beyond the Standard Model. The prospects
for the new few years are indeed exciting.